\documentclass{sigchi}

\usepackage{balance}       % to better equalize the last page
\usepackage{graphics}      % for EPS, load graphicx instead 
\usepackage[T1]{fontenc}   % for umlauts and other diaeresis
\usepackage{txfonts}
\usepackage{mathptmx}
\usepackage{color}
\usepackage{booktabs}
\usepackage{textcomp}
\usepackage{microtype}        % Improved Tracking and Kerning
\usepackage{ccicons}          % Cite your images correctly!
\usepackage{todonotes}

\usepackage[nodayofweek]{datetime}

\usepackage{floatrow}
\usepackage{wrapfig}
\usepackage{pifont}
\usepackage{layouts}
\def\plaintitle{Help, It Looks Confusing: GUI Task Automation Through Demonstration and Follow-up Questions}

\def\plainkeywords{Programming by Demonstration; GUI Automation; Action Segmentation and Recognition.}

\makeatletter
\def\url@leostyle{%
  \@ifundefined{selectfont}{
    \def\UrlFont{\sf}
  }{
    \def\UrlFont{\small\bf\ttfamily}
  }}
\makeatother
\def\pprw{8.5in}
\def\pprh{11in}

\definecolor{linkColor}{RGB}{6,125,233}
\def\ie{{\it i.e.,\ }}
\def\etc{{\it etc.,\ }}
\def\eg{{\it e.g.,\ }}
\def\etal{{et~al.}}
\def\vs{{\it vs.\ }}

\newcommand{\cmark}{\ding{51}}%
\newcommand{\xmark}{\ding{55}}%

\toappear{\scriptsize Permission to make digital or hard copies of all or part of this work for personal or classroom use is granted without fee provided that copies are not made or distributed for profit or commercial advantage and that copies bear this notice and the full citation on the first page. Copyrights for components of this work owned by others than ACM must be honored. Abstracting with credit is permitted. To copy otherwise, or republish, to post on servers or to redistribute to lists, requires prior specific permission and/or a fee. Request permissions from permissions@acm.org. \\
{\emph{IUI 2017, March 13--16, 2017, Limassol, Cyprus.} } \\
Copyright is held by the owner/author(s). Publication rights licensed to ACM. \\
ACM ISBN 978-1-4503-4348-0/17/03\ ...\$15.00.\\
http://dx.doi.org/10.1145/3025171.3025176}
\begin{document}

\CopyrightYear{2017} 
\setcopyright{acmlicensed}
\conferenceinfo{IUI 2017,}{March 13 - 16, 2017, Limassol, Cyprus}
\isbn{978-1-4503-4348-0/17/03}\acmPrice{\$15.00}
\doi{http://dx.doi.org/10.1145/3025171.3025176}
\title{\plaintitle}

\numberofauthors{1}
\author{
  \alignauthor{Thanapong Intharah, Daniyar Turmukhambetov, and Gabriel J. Brostow\\
  	\affaddr{Computer Science,}\\
    \affaddr{University College London,}\\
  	\affaddr{Gower Street,}\\
  	\affaddr{London, United Kingdom}\\
  	\email{\{t.intharah, daniyar.turmukhambetov.10, g.brostow\}@cs.ucl.ac.uk}\\}
%   }\alignauthor{Daniyar Turmukhambetov\\
%   	\affaddr{Computer Science,}\\
%     \affaddr{University College London,}\\
%   	\affaddr{Gower Street,}\\
%   	\affaddr{London, United Kingdom}\\
%   	\email{daniyar.turmukhambetov.10\\@ucl.ac.uk}\\
%   }\alignauthor{Gabriel J. Brostow\\
%   	\affaddr{Computer Science,}\\
%     \affaddr{University College London,}\\
%   	\affaddr{Gower Street,}\\
%   	\affaddr{London, United Kingdom}\\
%   	\email{g.brostow@cs.ucl.ac.uk}}\\
}

\maketitle

\begin{abstract}
Non-programming users should be able to create their own customized scripts to perform computer-based tasks for them, just by demonstrating to the machine how it's done. To that end, we develop a system prototype which learns-by-demonstration called HILC (Help, It Looks Confusing). Users train HILC to synthesize a task script by demonstrating the task, which produces the needed screenshots and their corresponding mouse-keyboard signals. After the demonstration, the user answers follow-up questions.
%Non-programming users should be able to train virtual robots to perform computer-based tasks for them. One would think that training in an all-digital noise-free environment should be easy. However, one-shot learning of a task is actually quite hard because so many mouse-clicks and key-presses are ambiguous. Also, recognizing individual actions in a task is not enough to reproduce the task, or to generalize it. For example, the intended reproduction of a copy-paste task could mean changing what is copied, or where it is pasted, or both.

We propose a user-in-the-loop framework that learns to generate scripts of actions performed on visible elements of graphical applications. While pure programming-by-demonstration is still unrealistic, we use quantitative and qualitative experiments to show that non-programming users are willing and effective at answering follow-up queries posed by our system. Our models of events and appearance are surprisingly simple, but are combined effectively to cope with varying amounts of supervision.

The best available baseline, Sikuli Slides, struggled with the majority of the tests in our user study experiments. The prototype with our proposed approach successfully helped users accomplish simple linear tasks, complicated tasks (monitoring, looping, and mixed), and tasks that span across multiple executables. Even when both systems could ultimately perform a task, ours was trained and refined by the user in less time. 
\end{abstract}

\category{H.5.2.}{Information Interfaces and Presentation}{User Interfaces} - {Graphical user interfaces (GUI)}


\section{Introduction}\label{sec:intro}
Millions of man-years are spent laboring in front of computers. Although some of that work is creative, many tasks are repetitive. Able-bodied users find it increasingly tedious to repeat a task tens or hundreds of times. Even more severely, users with special needs can find even a second or third repetition of a task to be prohibitively hard. We see computer vision in the domain of desktop and mobile Graphical User Interfaces (GUIs) as a delicate prerequisite, needed to make effective virtual personal assistants. 

The proposed approach examines the many small and non-obvious challenges to learning-by-demonstration in a GUI world. %Mobile phones and proprietary apps make the challenges of OS-event-sniffing insurmountable. 
While template-matching of icons and scripting of macros and bots are low-tech by modern vision standards, we start with these technologies because they are effective. The overall contributions of our approach are that: 
\begin{itemize}
%\begin{enumerate}
  \item Non-programming users can teach a task to our system, simply by demonstrating it, either on a computer with a sniffer program or through screencast-video.
  \item The system asks the human for help if parts of the demonstrated task were ambiguous.
  \item The same or other users can run the task repeatedly, giving them functionality that was previously hard to achieve or was missing entirely, \eg looping.
%\end{enumerate}
\end{itemize}

Tasks range from short single-click operations to inter-application chains-of-events. %A user may want the task to explicitly run once, \eg when the user requests it through speech recognition, or to loop over each item in a list, or to monitor for an alert or other visual cue. 
Our informal surveys revealed that each user had different tasks they wished to automate, but they agreed universally that the teaching of a task should not take much longer than performing the task itself. 

With the long-term aim of improving assistive technology, we separate the role played by the \emph{instructor} from that of the \emph{end-user}. For example, one person could use a mouse and keyboard (or an eye-tracker) to demonstrate a task and answer follow-on questions. Then the same or another person could run that learned task, \eg using speech. In this paper, we focus on algorithms to make the instructor effective. Moreover, we also split off the teacher role from the instructor role to support diverse demonstration inputs, \eg video tutorials \vs sniffer programs.

Through our designated scenarios, we explore three interesting issues of general Programming by Demonstration systems. For the linear tasks, we explore script generation; for the looping tasks, we explore generalization of the generated scripts; and for the monitoring tasks, we explore invocation of scripts. 

\textbf{Overview of Challenges}
At a high level, our system collects observations while the instructor performs a task, then it finds confusing looking patterns that call for the instructor to give more input. Once the task is learned and saved as a script, it can be called up by the end-user to run one or more times, or to monitor for some visual trigger before running.

A number of challenges make this an interesting technical problem that relates to object recognition, action recognition, and one-shot learning. The instructor's computer can be instrumented with a sniffer program, that records mouse/keyboard events, and screen-appearance. But, for example, a click-drag and a slow click are still hard to distinguish, and hugely varying time-scales make sniffed observations surprisingly hard to classify. We also explore learning of tasks from pre-recorded screencast videos, which display noisy details of key/mouse events. Further, clicking someplace like a File-menu usually means the task calls for that \emph{relative} location to be a target, but what if other parts of the screen have similar looking menus? The teacher can help find visual cues that serve as supporters. 

Also, while demonstrating, the instructor runs through a linear version of the task, and can indicate if some part of the chain of actions should actually become a loop. Template matching may reveal that the to-be-looped segment was applied to one unique target (an icon, a line of text, \etc), but the system can request further input to correctly detect the other valid targets on the screen. This functionality is especially useful when a looping task must ``step'' each time, operating on subsequent rows or columns, instead of repeatedly visiting the same part of the screen.

\section{Related Work}\label{sec:related}
The problem of learning-by-demonstration in a GUI world has challenges related broadly to four areas: action recognition from video, semantic understanding of GUI environments, analysis of video tutorials, and program synthesis.

\vspace{-6pt}\textbf{Joint Segmentation and Classification of Action:}
Hoai~\etal~\cite{Hoai2011} proposed an algorithm to jointly segment and classify human actions in longer video. Most other action-recognition works just classified pre-segmented clips. Like Hoai~\etal, we have unsegmented and long videos.
%They used dynamic programming to solve the problem by maximizing the different between the SVM score of the winning class and any other classes.
Shi~\etal~\cite{Shi2008,Shi2011} also segment and classify human actions. They use a Viterbi-like algorithm for inference, similar to ours, but they used Hamming distance to measure the loss between labels of two consecutive frames. Other action-detection methods tend to be slow and ill-suited for GUI problems.

\vspace{-6pt}\textbf{Semantic Understanding of GUI Environments:}
Dixon~\etal~\cite{Dixon2011} study models of GUI widgets, \eg buttons, tick boxes, text boxes, \etc and associate parts into hierarchical models from only visual information. This work and its extension~\cite{Dixon2014} aim to reverse-engineer GUI widgets to augment and enhance them. Hurst~\etal~\cite{Hurst2010} studied GUI targets the user can manipulate, for human performance assessment and software usability analysis. While having an accessibility API simplifies task-interpretation, it assumes the softwares's developers want to invest extra effort.
%This make the application of this work limited to certain applications.
Similarly, Chang~\etal~\cite{Chang2011} developed an accessibility API based system which also uses visual analysis to establish the hierarchical structure of the GUI. This also improved text analysis and processing. In contrast, we aim to recognize cross-application actions without access to software internals, mimicking what a human personal assistant would learn only from observation.

\vspace{-6pt}\textbf{Analysis of Video Tutorials:}
Grabler~\etal~\cite{Grabler2009} presented a system that generates a photo manipulation tutorial from an instructor's demonstration in the GIMP program. The system generates each step of the tutorial by accessing changes in the interface and the internal application state. The source code of GIMP was modified to allow such access. The authors also proposed preliminary work on transferring operations from the user demonstration to a new target. Again, this approach relies heavily on privileged access to one application. Chronicle~\cite{Grossman2010} is a system that allows users to explore the history of a graphical document, as it was created through multiple interactions. The system captures the relation between time, regions of interest in the document, tools, and actions. This rich information allows users to play back the video for a specific time or region of interest, to replicate the result or to understand the workflow.

The Pause-and-Play system~\cite{Pongnumkul2011} helps users better learn to use an application from a video tutorial. The system finds important events in the video tutorial, and links them with the target application. This allows the system to automatically pause the video by detecting events in the application, while the user is following the steps of the tutorial. However, the detection of events in an application is implemented again through the API of that application. The Waken system~\cite{Banovic2012} processes video tutorials using consecutive frame differences to locate the mouse cursor and the application widgets on which the video tutorial is focusing. Moreover, the system also infers basic widget actions executed by the user, and characterizes widget behavior. They  applied the system to a tutorial of a video player, allowing users to directly interact with the widgets in the application. However, this system mainly relies on pre-designed heuristic rules to detect cursors and widgets.

EverTutor~\cite{Wang2014} is a system that processes low-level touch events and screen-captured images on a smartphone to automatically generate a tutorial from a user demonstration, without binding itself to one application. A tutorial generated by that system can interactively guide the user through a process. Although this system approaches their problem in a similar way to our system for processing low-level mouse and keyboard input events and screen-captured images, their algorithmic details are omitted.

\vspace{-6pt}\textbf{Program Synthesis}
Our work is related to two sub-areas of program synthesis from user inputs: Programming by Demonstration and Programming by Example. Our system fits best in the Programming by Demonstration category. Programming by Demonstration systems (PbD) have the user perform a task as an input, and output a computer program that performs that task. In contrast, Programming by Example systems (PbE) learn the model from input-output pairs provided by a user, and generate a computer program that does the task.
Although research on Programming by Demonstration~\cite{lau2008pbd,cypher1993watch} and Programming by Examples~\cite{Gulwani16} shows promising results, the systems usually work within closed ecosystems where PbD can directly access software states and can manipulate the state. Automating user interaction with GUI objects, where working across different ecosystems is one of the most common scenarios, is still under-developed. The main contribution of our work is to make a PbD system be domain independent by relying on Computer Vision techniques rather than an Accessibility API.
%with ~\cite{Gulwani16}, work in GUI automating domain is still growing very slowly. 

Koala~\cite{Little2007} and CoScriptor~\cite{Leshed2008} built platforms to generate automation scripts and business processes with a loose programming approach, and to share the generated process across an organization. The aim was to personalize a process to a particular end-user's needs. Those systems focused only on web-based processes, since web-page source code is readily accessible and machine-readable.% However, it only supports web-based processes. %\tinote{Sloppy programming is pseudo-natural language instructions that created by the instructor and the system try to interpret and match with The Document Object Model (DOM) of the website.}

Sikuli~\cite{Yeh2009} was the first to provide programmers with an API to process and interact with desktop applications using computer vision techniques. This easy-to-use system can interact with multiple applications and can be run on multiple platforms. It was extended to many applications~\cite{Chang2010,Yeh2011}. However, the automation script has to be programmed by the instructor, hence, Sikuli Slides~\cite{Sikulislides} was developed to simplify the script generation process to allow less programming-skilled users to generate automation scripts. Instead of a coded script, Sikuli Slides represents a process as a Powerpoint presentation with annotations of basic actions on each of the slides. This presentation can be executed as a script.
Sikuli Slides also provides a recorder application, so a user records a process and saves it as a starting draft of the desired presentation. Our user study compares our framework and this baseline on linear tasks.

Sheepdog~\cite{lau2004sheepdog,Castelli201094} and Familiar~\cite{paynter2000automating} are PbD systems that focus on looping tasks. Our system deals with looping tasks through a different approach, so instead of requiring multiple demonstrations, we ask users to demonstrate once and then give examples of iterators.

\vspace{-6pt}\section{Learning Demonstrated Tasks}\label{sec:system}

Our system, HILC\footnote{Project page: \url{http://visual.cs.ucl.ac.uk/pubs/HILC/}}, has three phases: demonstration, teaching, and running. First, an instructor can choose to record their demonstration as either a screencast video (so highlighting mouse and key presses) or through custom-made sniffer software.
Both methods have pros and cons, which we discuss further in the Demonstration Phase section.
Next, during the teaching phase, the system performs joint segmentation and classification of basic actions. To generalize the observed actions, we introduce a human teacher (can be the same person as the instructor) who can help the system refine its pattern detectors. The challenges and proposed solutions of joint segmentation and classification, and for human in the loop training, are in the Teaching Phase section. %Our system workflow of the teaching phase can be found in Figure~\ref{fig:teaching}
Finally in the Running phase, the system performs actions according to the transcript generated by the instructor and improved by the teacher. Our system also generates a Sikuli-like script for visualization purposes. A stand-alone runtime script in pure python is made using the PyAutoGUI package. %The transcript is similar to a Sikuli~\cite{Yeh2009} script. %The details of this phase are provided in section \ref{sec:implementation}.

To start, we define a set of basic actions that our system can perform, and our system needs to jointly segment and recognize these actions from the input demonstration. The set of basic actions is composed of Left Click, Right Click, Double Click and Click Drag. 
%In section Recording the Task, we will describe recording phase in detail.
%Teaching phase is covered in section Classification and Segmentation of Basic Actions and section Interactive Training of Pattern Detectors with Few Examples.
%Detail of applying phase is postponed to the Implementation section.

\vspace{-6pt}\section{Demonstration Phase}
In this phase, an instructor who knows how to complete the task demonstrates the task while only recording video or while running the sniffer program. If the user chooses to record the task as a video, the system will pre-process the video and create the unified format log-file from every frame of the video. On the other hand, if the user chooses to record the task via a sniffer, the system will record every signal produced by the user to the unified format log-file. %when user choose to record the task with the sniffer  

\vspace{-6pt}\subsection{Recording the Task}\label{sub:recording}
\begin{figure}[ht]
{\caption{Example of a log-file which merges our inputs from both video and sniffer data.}\label{fig:log}
}
{\includegraphics[width=8cm]{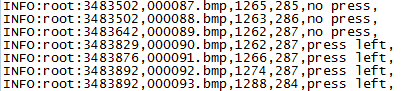}}
\end{figure}

For both input methods (video-only or sniffer) we define a unified sensor-data format. Functioning like a log-file (see Figure~\ref{fig:log}), each entry records the screenshot image and the low-level status of the machine: its mouse button status, mouse cursor location, and the keyboard's key press status. %show format of log file
A sniffer can access useful information directly from the OS, but can't determine the class of the basic action. It does naturally generate a log-file from the demonstration. However, one must take into account time intervals between each log entry, inconsistent machine lag, and storage space of screenshots. Detaila are presented in the Implementation section.

To help the system recognize whether a demonstrated segment is a linear, looping, or monitoring/standby task, we use special key combinations that the user was briefed on before using the system. We use the terms monitoring and standby interchangeably and they both refer to the same kind of task. The details of the recording step and the special key combination are presented in the User Study Scenarios section.

Although processing a video-only demonstration is slower, it is more versatile because end-users can leverage  pre-recorded and internet-shared videos as an input demonstrations. %\tinote{this method is free from trusted software issues.} 
However, processing of demonstration videos has many challenges, such as locating the mouse cursor, retrieving low-level mouse/keyboard events from screencast messages, removing the mouse pointer from the video's screenshot images (\ie capturing a template of the button without the cursor's occlusion), and noise/compression in the video recording.

We describe the implementation details of our approach for coping with the aforementioned issues for both sniffer and video inputs in the next section.%\ref{sec:implementation}.

\vspace{-6pt}\subsection{Implementation}\label{sec:implementation}
%In this section we describe implementation details of the Sniffer, Looping signal, Standby signal, Teaching phase, Running phase and Video Processing. 
The system was implemented in Python~2.7 and it was deployed on Microsoft Windows 7 64-bit machines with Intel Core i5-3317u @1.70GHz CPU and 8GB of RAM. 

We implemented a custom sniffer which records low-level mouse and keyboard events. Log-file entries and screenshots were only saved before and after each of the mouse and keyboard status changes, to keep the hard drive i/o from slowing down the machine. We also re-sample records to have log-files where the time difference between records is $1000/30$ milliseconds, to make sniffer log-files versions agree with video versions that sample screenshots at $30$ frames per second.
%Please note that the sniffer version only works on Windows Operating System because of the using of pywin32 package.
%The sniffer only saves screenshots before and at the time the mouse buttons is pressing and releasing to avoiding the lagging of the system from the sniffer software.
%Moreover, we only record a log-file entry when there is a mouse or keyboard actions. %this is to one important issue: un-even log entries which is different from the log-file that generate from the video. To make the log-files from both input method the same standard, we pre-process log-files from the sniffer by enforcing the log-files have exactly 30 entries/second as same as video input 30 frames/second.

Using an existing video as a demonstration requires that we address two important issues: how can the system retrieve low-level mouse and keyboard status information, and how can the system remove the mouse cursor from video frames for clean pattern extraction?

Video tutorials are commonly recorded with specialized screencasting software that renders visual indicators of mouse events, left-right button pressing, and keystrokes. Hence, we assume that the demonstration video was recorded with key-casting software. Each key-caster is different, but we trained for KeyCastOW for Windows and Key-mon for Unix. Others could easily be added. Thus, we extract key-cast display locations from the video and recognize the information for each frame via the OCR software Tessarract. The mouse status information displayed by the key-cast software is still low-level, similar to sniffer output.%\ref{tessarract}. 
The mouse cursor location is with by a Normalized Cross Correlation detector of the mouse pointer template. Videos of demonstrations have an inherent problem of the mouse cursor occluding a target pattern during a basic action. To overcome this, we detect significant appearance changes of the screen and use a temporal median filter to remove the cursor to create a clean mouse-free screenshot.

In the Demonstration Phase, the instructor's basic actions are recorded by the system.
Hence, to interact with the system, for example, to start or stop the recording, the instructor triggers explicit signals. In our implementation, these are special key combinations. The detection of loops or standby patterns is not automatic, and has to be flagged by the instructor. Therefore, we define three special key combinations for the different signals: End of Recording, Looping, and Standby, which are discussed in detail before we explain each designated scenario in the Evaluation and Results section. %Further, the teaching phase also has special key combinations for specifying true and false detections, supporters, \etc.

%At demonstration phase, the instructor can indicate the sequences of basic actions where the user want the Looping or Standby to be applied by pressing combinations of keys (\eg Ctrl+Shift+l, Ctrl+Shift+Pause to indicate the start of the looping, the end of the looping%, and the end of instances selection
%, respectively).
%Then the instructor indicates the end instance of instances selection with the looping signal.
%So, the instructor specifies the trigger pattern with a special key combination. Then, the instructor performs a sequence of basic actions which are to be performed after detection of the trigger pattern.
%Likewise, we define the key combinations for Standby as Ctrl+Shift+w or Ctrl+Shift+PrtScr.

%We implement the interactive module that allow the system to query from the human user using Pygame module. 
%Please refer to supplementary materials for the system workflows that show cases in which the system queries the teacher.

\vspace{-6pt}\section{Teaching Phase}
\begin{figure}
%\floatbox[{\capbeside\thisfloatsetup{capbesideposition={right,top},capbesidewidth=4cm}}]{figure}[\FBwidth]
{\includegraphics[height=13cm]{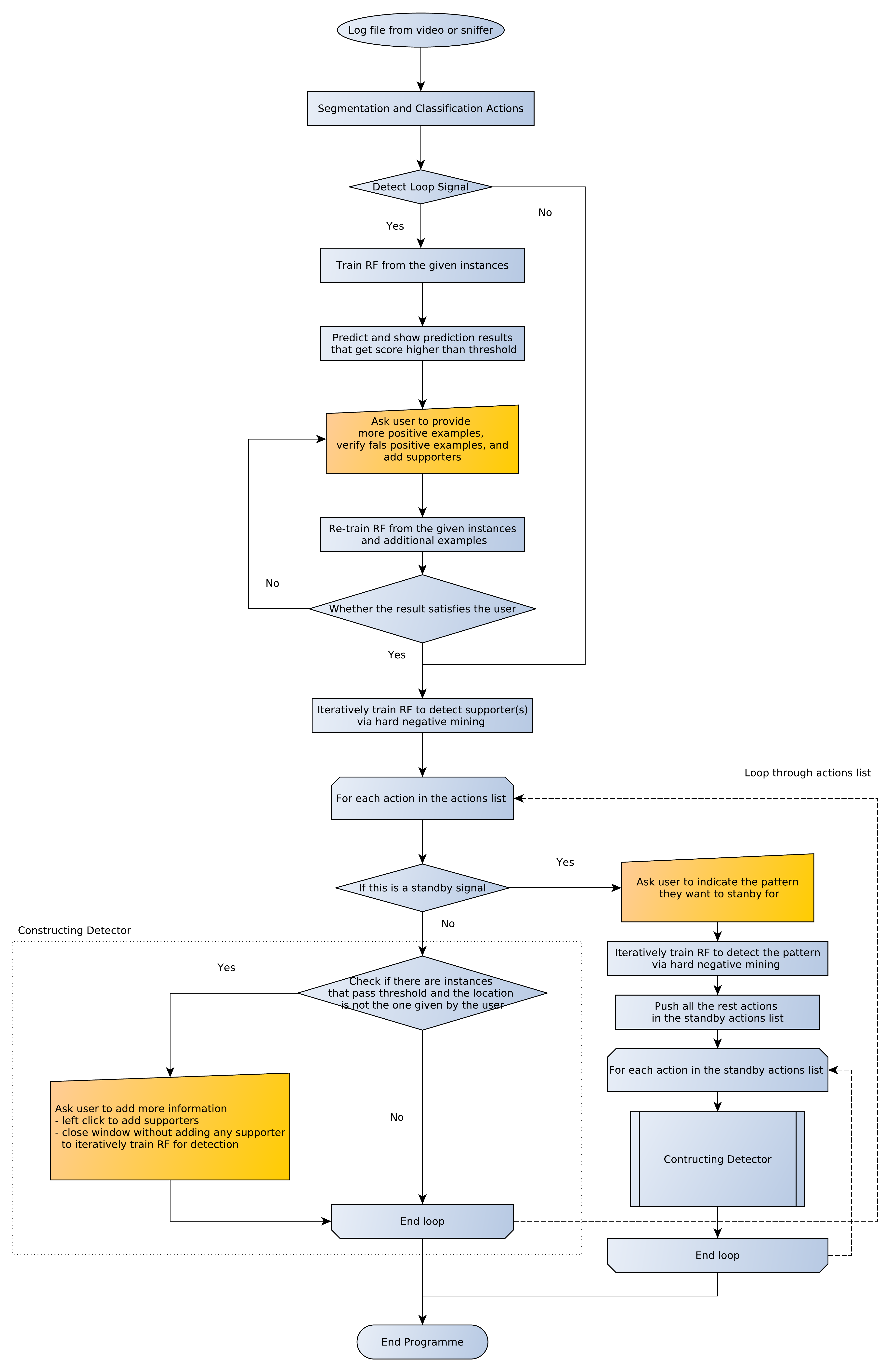}}
{\caption{Our system workflow for the teaching phase. The yellow boxes indicate where the system poses questions to the teacher.}\label{fig:teaching}}
\end{figure}

This phase of the system takes as input the log-file from the Demonstration Phase, and then produces a transcribed script that consists of a sequence of basic actions. The system workflow of the teaching phase is in Figure~\ref{fig:teaching}.

\vspace{-6pt}\subsection{Classification and Segmentation of Basic Actions}\label{sub:classandseg}
%Classification and Segmentation basic actions from video and sniffer (use common input). pros, cons, and problem of both methods.\\
Although the log-file contains all of the low-level information, identifying basic actions is not straightforward. For example, the ``left mouse button pressed'' status can be present in the log-file at multiple consecutive entries for a single ``Left Click'' basic action. Further, it is ambiguous if a left click is a single ``Left Click'' basic action, or a part of thw ``Double Click'' basic action. Moreover, variability based on how individual users interact with the computer via peripheral devices makes it hard for deterministic rules to distinguish between different basic actions.%, see training system flow. %\ref..
This drives the need for training data, though each user provides only very little. 

To create a system that can cope with ambiguities in recognizing basic actions, we treat the problem as a Viterbi path decoding problem: based on dynamic programming, our algorithm segments and classifies the basic actions concurrently. 

Let $Y = \{y_1,...,y_z\}$ be the set of all possible states, with a label for each temporal segment. $x$ is the observation of a segment, a feature vector representing part of a basic action.  The unary terms $U(y|x)$ are probability distribution functions over parts of basic actions, learned from pre-collected training data, The pairwise terms $P(\tilde{y},y)$ are the constraints that force consecutive parts of actions to be assigned to the same basic action. 

Let us define $A_k$ as a basic action made from a sequence of parts of the action $k$, $A_k = \{y_1^k,y_2^k,...,y_l^k\}\;;\;|A_k| = l$ and $y_l^k$ is the last part of $A_k$. $y^k_n$ is a part of the basic action $k$ where $n$ indicates a status change frame (key frame) in the log-file, \eg a frame where the mouse button status is changed, from press to release or vice versa. Pairwise term is defined as in Eq~\ref{eq:pairwise}:
\begin{equation}\label{eq:pairwise}
P(y_{v-1}^i,y_{v}^j) = \begin{cases}
    +1       & \quad \text{if } i=j \text{ and } y_{v-1}^i \text{ follows } y_{v}^j\\
    \;\;\;0  & \quad \text{if } i\neq~j \text{ and } y_v^i \text{ is the last}\\& \quad\text{part of } A_i\\
    -1  & \quad \text{otherwise}.\\
  \end{cases}
\end{equation}

%\tinote{To be clear, this is different from purely classification which attempts to classify well trimmed unknown actions, or purely segmentation which attempts to segment an action class from background action. Unlike either solely recognition or segmentation which attempts to recognize trimmed unknown action or segment an action class from background action.%positive action from negative action.} unary terms \ie probability distribution functions over basic actions,
\textbf{Learning Probability Distribution Functions}
One of the challenges of basic action classification is that each basic action has a different duration. For example, a ``Left Click'' usually spans a few milliseconds in the log-file, but ``Click Drag'' may span seconds. Moreover, each basic action also has its inter-variability in terms of action duration. %Training a classifier with features extracted from a time interval corresponding to the longest basic action would need multi-label classification. %\dtnote{need a clear example?}

Our approach is to train a Hinge-loss Linear SVM for each basic action using hard negative mining method, where each SVM may make a prediction on a different time interval. For an input time interval, we construct a feature vector by computing histograms of frequencies of different encoded records from the log-file. Each record is encoded by $3$ binary low-level states: the status of each of the mouse buttons, pressing the left button/ the right button, and whether the mouse is moving. The feature vector also encodes context information of each action by adding a histogram of a small time interval after an action is done. Finally, our feature vector for the Linear SVM has $4\times 2^3$ dimensions, the number $4$ is from $3$ histograms of equally divided time interval of an action, note that this is different from a part of an action $y^k_n$, plus $1$ context histogram. 

The prediction scores of SVMs are not proportional to each other, so must be scaled before use within the Viterbi algorithm. Hence, for each basic action, we train a Random Forest (RF), that inputs the prediction scores of SVMs as features and outputs the probability of each basic action class.

For any unknown time interval, we can construct a unary matrix which maps between a part of action to its probability distribution from trained RF for that action. The unary matrix $U_{s\times~c}$ has the shape (number of states$\times$number key frames in the input sequences). The number of states is defined as Eq~\ref{eq:nostates}
\begin{equation}\label{eq:nostates}
\text{number of states}=\sum_{A_k\in B}|A_k| \;,
\end{equation}
Please note that, we assign the same probability distribution to all parts of each basic action.
%Because each SVM classifier outputs unnormalized score based on distance from its separating hyperplane; so ranges are varied by data. 
%normalize scores of a class that hard to distinguish, negative and positive land near separating hyper plane, and a class that very to distinguish, negative and positive land far from the separating hyperplane

%\dtnote{ I'm not sure we need this: In addition, we learn basic actions' important statistics: minimum length, maximum length and the number of change signals of the action. The statistics are used for filtering action before passing it through the classifier. We also detect status change frames in the log-file as frames where the mouse button status is changed, from press to release or vice versa. These status change frames are used as possible starting and ending points of basic actions.}

\textbf{Annotation of Log-files for Basic Actions}
To annotate an unknown log-file, the system detects status change frames in the log-file and use them as key frames. Records in the log-file are then grouped into $N$ different time intervals indexed by the key frames and passed to the trained Random Forests to construct the unary matrix $U$. Lastly, we do inference for $\mathbb{Y}^*$, which is the sequence of $y^k_n$ that maximizes Eq~\ref{eq:viterbi} using the Viterbi dynamic programming algorithm, where
\begin{equation}\label{eq:viterbi}
\mathbb{Y}^*=\operatorname*{arg\,max}_{\mathbb{Y}^*} \sum_{v=1}^N (P(y_{v-1}^i,y_{v}^j) + U(y^j_v|x_v))
\end{equation}

\vspace{-2pt}\subsection{Interactive Training of Pattern Detectors with Few Examples}\label{sub:generalization}
%\section{Generalization and Interactive Training}\label{sec:generalization}
%In this sections, we list challenges of each part and present techniques how we overcome those challenge.
%Generalization and interactive training.
To perform any basic action, the system needs to learn the appearance of the target of the basic action. For linear tasks, the system needs to find a single correct location of the target pattern. However, for looping tasks, the system has to find multiple correct locations of the targets of the looping task, and thus the system needs to generalize about the target pattern appearance. In both cases, the target pattern may have appearance variations. For example the icon of the file may have moved on the desktop and has a different section of the wallpaper as its background.

\begin{figure*}[ht]
\begin{tabular}{cc}
\bmvaHangBox{\fbox{\includegraphics[height=1.5cm]{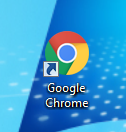}}}&
\bmvaHangBox{\fbox{\includegraphics[height=1.5cm]{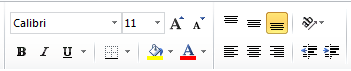}}}\\
(a)&(b)\\
\bmvaHangBox{\fbox{\includegraphics[width=4.2cm]{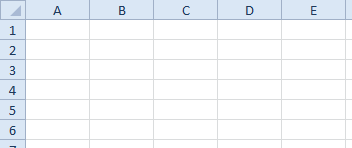}}}&
\bmvaHangBox{\fbox{\includegraphics[width=4.2cm]{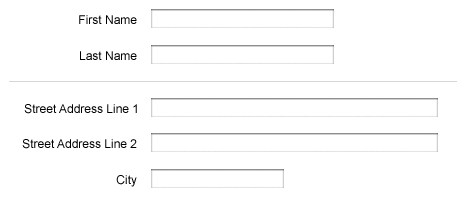}}}\\
(c)&(d)
\end{tabular}%\vspace{-12pt}
\caption{Target patterns in (a) \& (b) are distinguishable on their own. The spreadsheet cells in (c) need row and column names to differentiate between each other. Text fields in registration forms in (d) can be distinguished by the text field labels. Supporter help distinguish locally ambiguous patterns.}
\label{fig:pattern}
\end{figure*}

Target patterns can also be categorized into two groups: patterns that are discriminative on their own and patterns that need extra information to be distinguishable, see Figure~\ref{fig:pattern}. Hence, the system needs to treat each of these possibilities differently and we also introduce a concept of supporters here.

\textbf{Supporters}
The supporters are salience patterns that have certain offsets to the target pattern. 
In linear tasks, a spreadsheet program's row and column names distinguish similar-looking cells, \eg Figure~\ref{fig:pattern}(c); or field names distinguish similar looking textboxes, Figure~\ref{fig:pattern}(d).
In the Running Phase, the system uses the same technique that is used for detecting the target pattern, to detect supporters. %NCC to detect the patterns and supporters.
The target patterns give votes to each detection location, but supporters give votes to the offset locations.

\begin{figure}[ht]
%\floatbox[{\capbeside\thisfloatsetup{capbesideposition={right,top},capbesidewidth=5cm}}]{figure}[\FBwidth]
{\caption{An example of a supporter for a looping task. The table shows names of characters and actors/actresses of a popular TV show. The names in each column have similar appearance, so, if the user intended to loop through one of the columns, marking the column name as a supporter will help the system to distinguish between columns.}\label{fig:loopsup}
}%from https://en.wikipedia.org/wiki/List_of_Game_of_Thrones_characters
{\includegraphics[width=5cm]{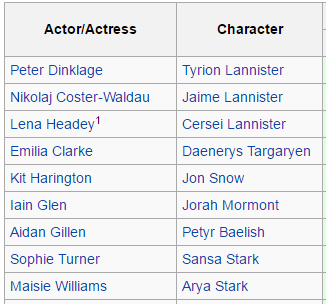}}
\end{figure}

\begin{figure*}[ht]
\begin{tabular}{cc}
\bmvaHangBox{\fbox{\includegraphics[height=3.2cm]{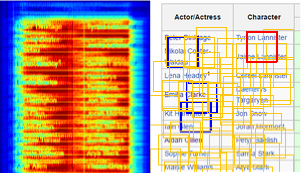}}}&
\bmvaHangBox{\fbox{\includegraphics[height=3.2cm]{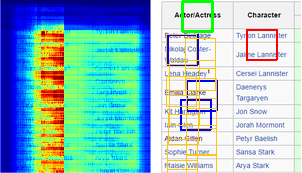}}}\\
(a)&(b)\\
\end{tabular}\vspace{-6pt}%\vspace{-12pt}
\caption{An example of a spatial supporter. Blue boxes are user provided positive examples, red are user provided negative examples, yellow are target detections, and a green box indicates a user-provided supporter. (a) and (b) demonstrate detection performance without and with a spatial supporter; red means a high detection score. In (a), the left image shows the heatmap of target detection scores, and right shows detected targets. In (b), the left image shows the heatmap of target detection and spatial supporter scores, and right shows detected targets. The spatial supporter successfully suppressed all similar looking patterns under the Character column in heatmap (b) so that the system is able to detect only desired targets under the Actor/Actress column}
\label{fig:loopsup_results}
\end{figure*}

Supporters for looping tasks work differently from the supporters for linear tasks, as a fixed offset is not informative for multiple looping targets. Hence, the supporters for looping provide x-axis and y-axis offsets to the targets, see Figure~\ref{fig:loopsup}. The final result is the average of target pattern detectors and spatial supporters, see Figure~\ref{fig:loopsup_results}.
%\dtnote{Don't need this: Our PbD system must handle two distinct learning scenarios: learning to perform linear tasks and learning looping tasks.  The two scenarios are different in their goals.  Learning for linear tasks requires the system to find one exact pattern and generalize about the background of the target pattern. Whereas the latter requires the system to be able to generalize about both the pattern itself and the background.}

\textbf{Follow-up Questions:} For different kinds of tasks, the system asks for help from the teacher differently.
\newline\textbf{Linear tasks}
Every basic action of linear tasks has a unique target. The linear tasks can be executed multiple times, but each run performs the same task on the same unique targets. 
For each basic action of the linear task, the system has to learn the corresponding target pattern from only one positive example.
To train the detector for the target pattern, the system initially performs Normalized Cross Correlation (NCC) matching with the given positive example on the screenshot image when the basic action was about to be executed, to prevent the pattern from changing appearance after the action executed. 
If there are false-positive locations with high NCC score, the system asks the teacher to help the system to distinguish between true and false positive examples by providing a supporter(s). %The supporters, are similar to target patterns, but instead of voting on the detected location, they vote for a location at a fixed location offset. 
After teachers provide supporters, the system uses NCC as the detector for both the target pattern and the supporters.
If the teacher does not provide any supporters, the system assumes that this pattern is distinguishable on its own. More false positive patches are mined to retrain RF until the system correctly detects the target location using raw RGB pixel values of every position in the patch as features, until the system can distinguish between true positive and false positive patches.

\vspace{-6pt}\textbf{Looping tasks}
Looping tasks are a generalization of linear tasks, so that in the running phase, each run of the task iterates over a set of targets. For example, a linear task always prints a unique PDF file in a folder, but a looping task prints all PDF files in a folder by looping over each PDF file icon. 

For looping tasks, the instructor shows the task once on a single looping target, but the system needs to repeatedly perform the task on all looping targets that are similar to the pattern the instructor, or the teacher, or the end-user had specified.
In the Demonstration Phase, the instructor is asked to show more than one example of a looping target after demonstrating one complete iteration of a task. %allowed to show more than one example of the looping targets. After the end of the looping, the system requires the instructor to provide a few positive example of the target patterns by Ctrl+clicking at each example to help the system generalize about appearance of the target.
In the Teaching Phase, the system trains a RF with the provided positive examples, and random patches as negative examples.
Next, the system validates the RF predictions by asking the teacher to verify predicted positive and negative examples, and/or add supporters.%mark false positive detections, adding more positive examples, and adding supporters.

\vspace{-6pt}\textbf{Monitoring tasks}
In monitoring tasks, the system at the Running Phase perpetually runs in standby, looking for a specified visual pattern, to invoke the rest of the script. In the Demonstration Phase, the instructor indicates when the invocation pattern appears, then demonstrates the task itself. In the Teaching Phase, the system asks the teacher to indicate which pattern needs to be detected. 

%\dtnote{ not needed: \small{[Image source:\url{https://en.wikipedia.org/wiki/List_of_Game_of_Thrones_characters}]}}

\vspace{-6pt}\section{Running Phase}\label{sec:running}
The main reason we separate the running phase from the demonstration phase and the teaching phase is that our ultimate goal for the system is to help end-users who are non-regular computer users, and disabled users, to complete tasks that might be hard for them but easy for others. The separated system is easy to execute by voice command or any other kind of triggering methods.

\begin{figure}[h]
%\floatbox[{\capbeside\thisfloatsetup{capbesideposition={right,top},capbesidewidth=4cm}}]{figure}[\FBwidth]
{\caption{Our system workflow for the running phase. The yellow box indicates user interaction.}\label{fig:running}}
{\includegraphics[height=7cm]{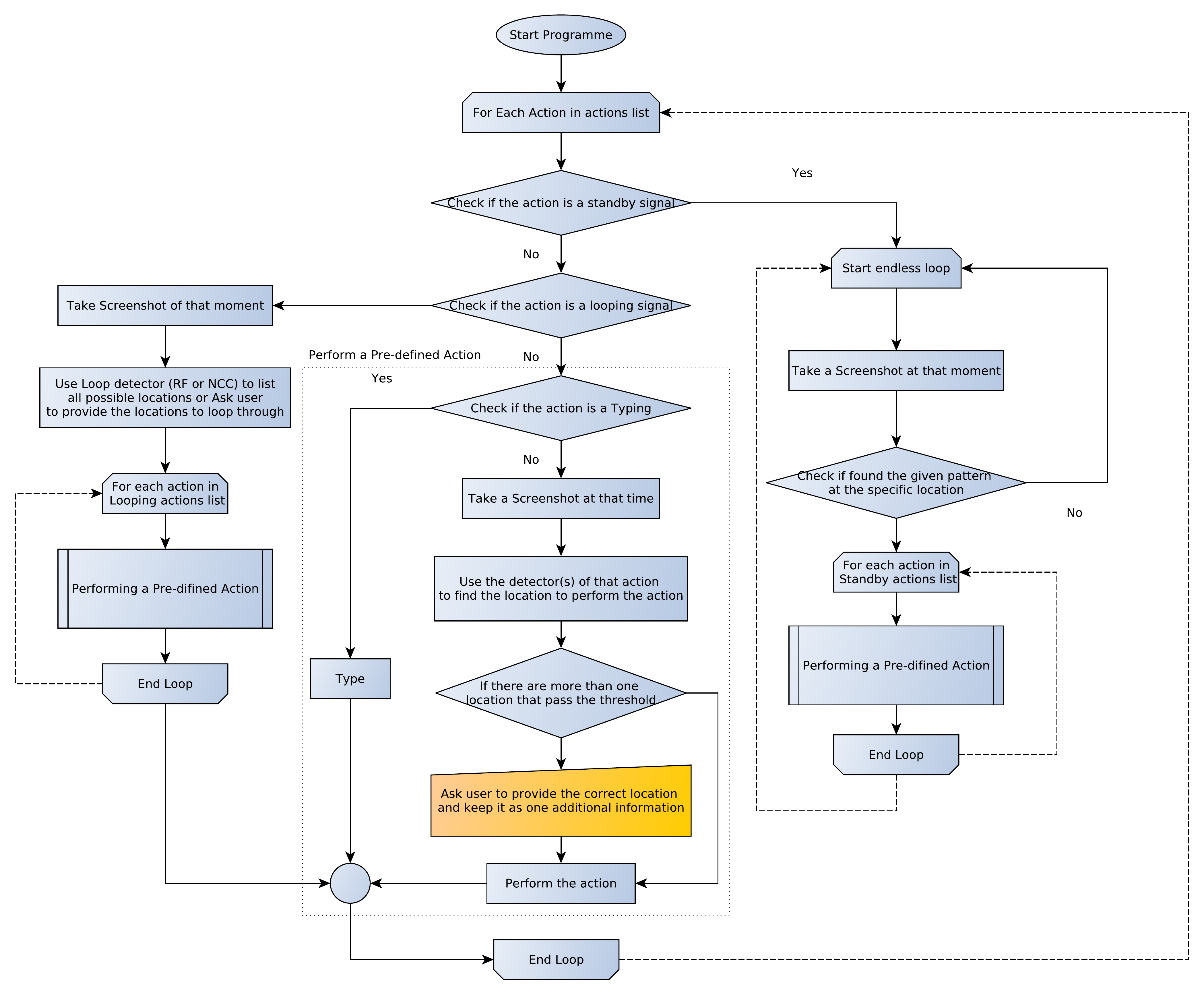}}\vspace{-6pt}
\end{figure}

In the running phase, the system sequentially executes each action of the interpreted sequence of actions from the teaching phase. When a special signal like Looping or Standby is found, the system executes the specific module for each signal. The Running phase's system flow is shown in Figure~\ref{fig:running}. For each normal basic action, the system starts by taking a screenshot of the current desktop and then looks for the target pattern and supporters (if available) using the trained detector(s). For the looping part, after taking a screenshot of the current desktop, the system evaluates every position on the screen with the trained detector RF, and applies the spatial voting from the supporter(s). After that, the non-maxima suppression and thresholding are applied respectively to the result, to get the list of positions to loop over. For the standby task, the system continuously takes a screenshot of a current desktop and evaluates the specified positions whether the target pattern may appear. When the target pattern is found, the system triggers the sequence of actions that the instructor designed, and then proceeds to the standby loop again.    

\section{Evaluation and Results}\label{sec:evaluation}
We evaluated our algorithm quantitatively through a small user study, and qualitatively to probe our system's functionality. Here, we document just nine use cases, and video of some of these is in the Supplemental Material. The only viable baseline PbD system available for comparison is Sikuli Slides\cite{Sikulislides}, because it too assumes users are non-programmers, and it too has sniffer-like access to user events. We collected (through a survey) and prioritized a larger list of scenarios for which our colleagues would like a virtual personal assistant to complete a task. We picked scenarios that spanned the different basic actions, and sampled the space between having just two steps and up to $17$ steps. Here we list the three scenarios tested in the seven-person user-study, and show its results in Table~\ref{tab:result}. Just these three linear tasks were picked because Sikuli Slides can not handle looping or standby tasks. Further scenarios are discussed in the qualitative evaluation.

The task in each scenario was assessed in terms of 1) transcription accuracy (evaluating classification and segmentation),  2) task reproduction, \ie measuring pattern detection generalization, and 3) time users took to demonstrate and then refine the task model.
In our studies, four of the participants had no programming exposure, two had taken a school-level course, and one was a trained programmer. Participants needed  1.5 - 2 hours because each completed all three tasks under both systems: they randomly started with either ours or Sikuli Slides, and then repeated the same task with the other system before proceeding to the next task. Before using both systems the users were briefed about the goal of the study as well as how to for using both systems for 20 minutes. In addition, the users were shown the videos of the instruction phase for each task before performing the task to ensure the users understand what are the tasks.   %\ttnote{Thanapong, please describe that users received training for using each system - how long or how presented to them.}
%Table \ref{tab:result} summarizes the user study.

\begin{figure}
\floatbox[{\capbeside\thisfloatsetup{capbesideposition={right,center},capbesidewidth=5cm}}]{figure}[\FBwidth]
{\caption{The instructor clicked on the speaker in the green box, but the system also detected a similar pattern - the speaker in the red box. In this situation, our system asks the teacher for a supporter(s), the yellow box, to  help with detecting the intended pattern.}\label{fig:scene1}}
{\includegraphics[height=2.7cm]{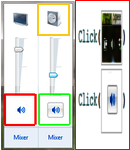}}\vspace{-6pt}
\end{figure}

\noindent\textbf{User Study Scenarios:} We evaluate the first three basic scenarios (linear tasks) quantitatively against Sikuli Slides and evaluate monitoring and looping tasks qualitatively. 
\newline\textbf{Linear:} Linear tasks are simply linear sequences of actions. They are the most basic type of task, and run only once. To record and edit all kinds of tasks, two common steps need to be done: First, at the end of a task-demonstration, the user presses the special keys combination Shift+Esc, to indicate the end of sequence. Second, also in the teaching phase, if the system cannot clearly distinguish between an input pattern and the other on-screen content, the system asks the teacher to click on supporter(s) near that pattern. 
% These scenarios, and especially $\#4-\#9$, are explained in greater detail in the supplementary material.
%\tinote{More detail about the scenario can be found in the supplementary materail \url{https://www.overleaf.com/5191107tqfxkr}}\\
\newline\textbf{1. Mute audio playback} (Linear) This simple and short task was actually non-trivial because the speaker icon in some Windows installations is not unique (Figure~\ref{fig:scene1}). Half the users had to refine the model, which for our system meant adding a supporter. All users produced a working model of this task using our system, and some users were able to produce a working model of the task using Sikuli Slides (the right speaker icon was correctly selected by chance).

\begin{figure}[h]
%\floatbox[{\capbeside\thisfloatsetup{capbesideposition={right,top},capbesidewidth=7cm}}]{figure}[\FBwidth]
{\caption{High-Contrast-Mode comparing with Normal Mode and (in the red box) the transcribed steps of the task demonstrated by our system.}\label{fig:scene2}}
{\includegraphics[height=4cm]{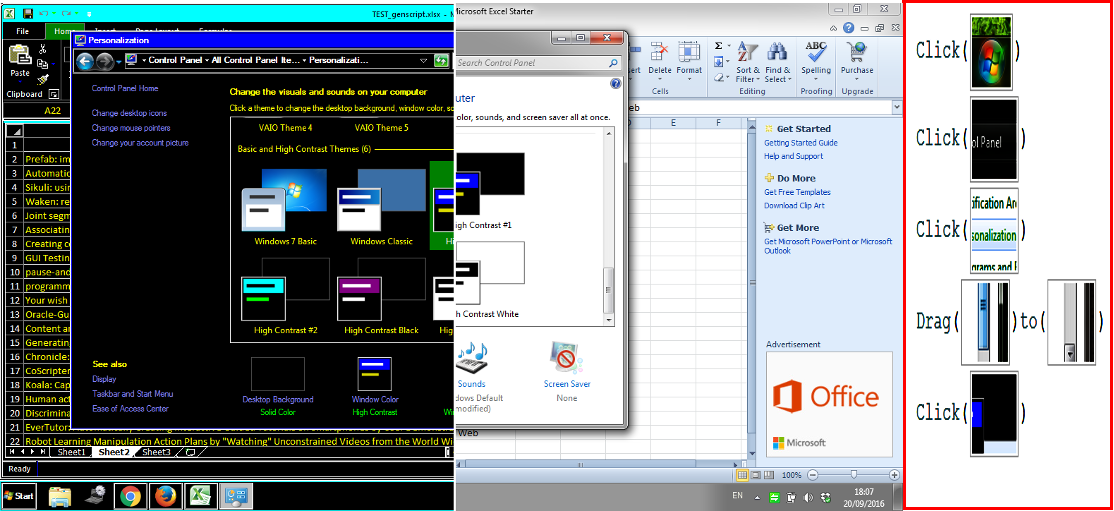}}\vspace{-8pt}
\end{figure}

\vspace{-6pt}\textbf{2. Turn on High-Contrast-Mode} (Linear) Some visually impaired \emph{end-users} may want to trigger this task through a speech recognition system. %Some operating systems offer such functionality directly. 
Here, our sighted \emph{instructors} were ultimately successful using both systems, but the study-supervisor had to walk users of Sikuli Slides through the extra steps of re-demonstrating the task and making and editing of screen-shots to refine that model.
This task involves the Click Drag action, which Sikuli Slides was never able to recognize when transcribing. Figure~\ref{fig:scene2} illustrates the High-Contrast vs default desktop modes, and our synthesized script for switching. Please note that High-Contrast-Mode also modifies the scale of objects on the screen.

\begin{figure}[h]
%\floatbox[{\capbeside\thisfloatsetup{capbesideposition={right,top},capbesidewidth=7cm}}]{figure}[\FBwidth]
{\caption{Steps to complete remote access via TeamViewer. Red lines link related patterns on the screen with the pattern in the transcript. Please note that performing the basic action DragTo from and to the same pattern has a similar effect as performing the basic action Click on that pattern. %Participants often unintentionally perform the basic action Dragto instead of the basic action Click. 
Our system is robust to this type of different-but-interchangeable action.}\label{fig:scene3}}
{\includegraphics[height=4cm]{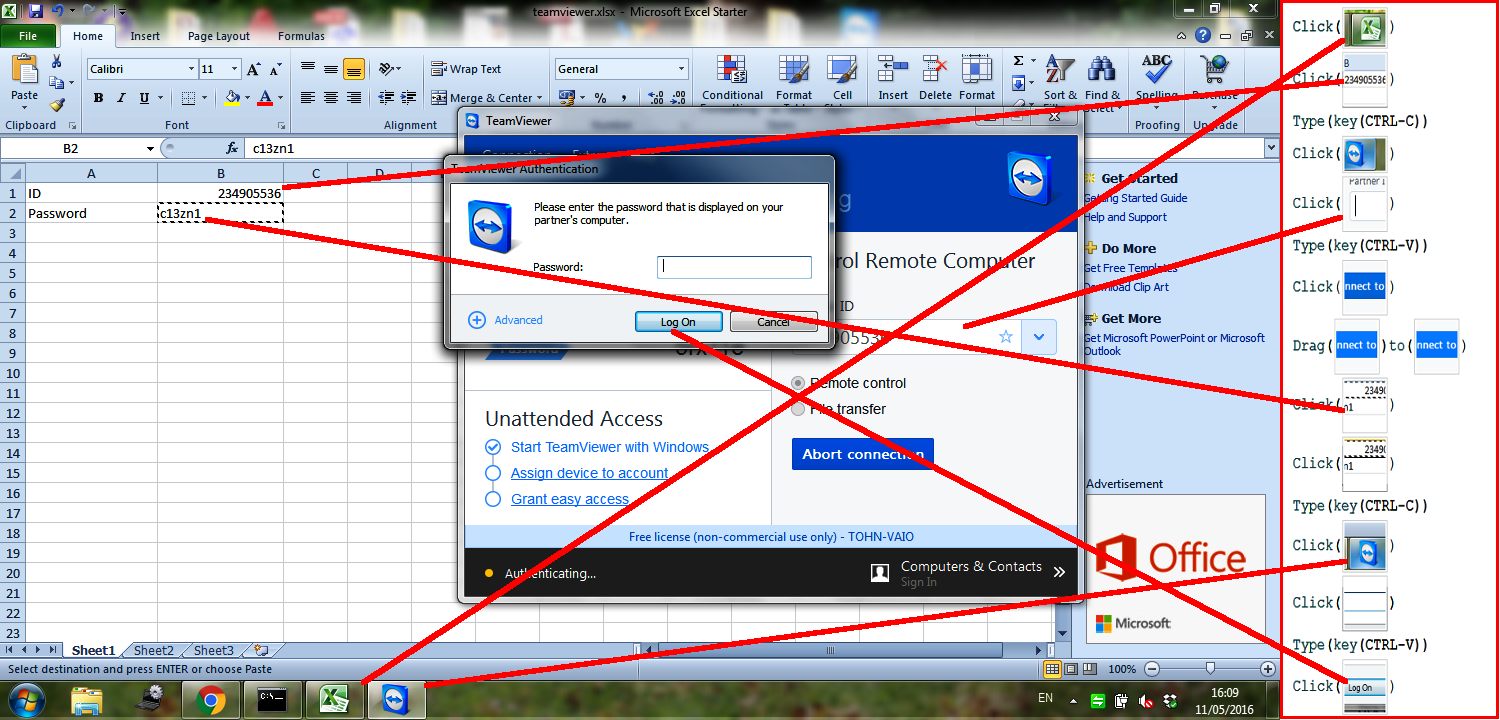}}\vspace{-8pt}
\end{figure}

\vspace{-6pt}\textbf{3. Remote access with TeamViewer} (Linear) This sysadmin (or mobile-phone testing) task consists of running the TeamViewer application and logging into another device, using an ID and password provided in a spreadsheet file. The teaching phase of the task involves helping the system clarify ambiguous patterns by adding supporters. We illustrate the task and the transcribed steps in Figure.~\ref{fig:scene3}.
%The task show the robustness of our system that can vary input texts via spreadsheet file or text file.
Inadvertently, this task proved impossible for Sikuli Slides users because it involves copy-pasting text, which that system does not capture keys combination shortcut. One user invented an alternate version of this scenario ($3.2$) where she tried to right-click and use a context menu to copy and paste, but we then realized that right-clicks are also not captured by Sikuli Slides. 

\noindent\textbf{Qualitative Evaluation: } The remaining scenarios can not be addressed using Sikuli Slides because they entail monitoring or looping tasks. We outline our new capabilities here, along with qualitative findings, and task illustrations, to better gauge success.
\newline\textbf{Monitoring:} Monitoring tasks run perpetually and then respond to specific patterns. %Right after started the script, our virtual robot keeps evaluating patterns from a specific location. 
When the specified pattern is detected, the script triggers the sequence of predefined actions. 

In the Demonstration Phase, instructors press the special key combination (standby signal), Ctrl+Shift+w or Ctrl+Shift+PrtScr, to indicate that the pattern, which we want the system to detect, has appeared. The instructor then performs a desired sequence of actions, such as a linear task. There is an extra step in the Teaching Phase, where the system asks the teacher to indicate where the invocation pattern \emph{can} occur (\eg anywhere, or in the taskbar). 
%In the teaching phase, There is an extra step which the system will ask a teacher to indicate where is the pattern which is needed to be detected. 

\begin{figure}[h]
%\floatbox[{\capbeside\thisfloatsetup{capbesideposition={right,top},capbesidewidth=4cm}}]{figure}[\FBwidth]
{\caption{YouTube Skip Ad. These advertisements show before or during a playing video for varying periods of times, and our system successfully closes them in Scenario 4, as soon as the text appears.}\label{fig:scene4}}
{\includegraphics[height=2.7cm]{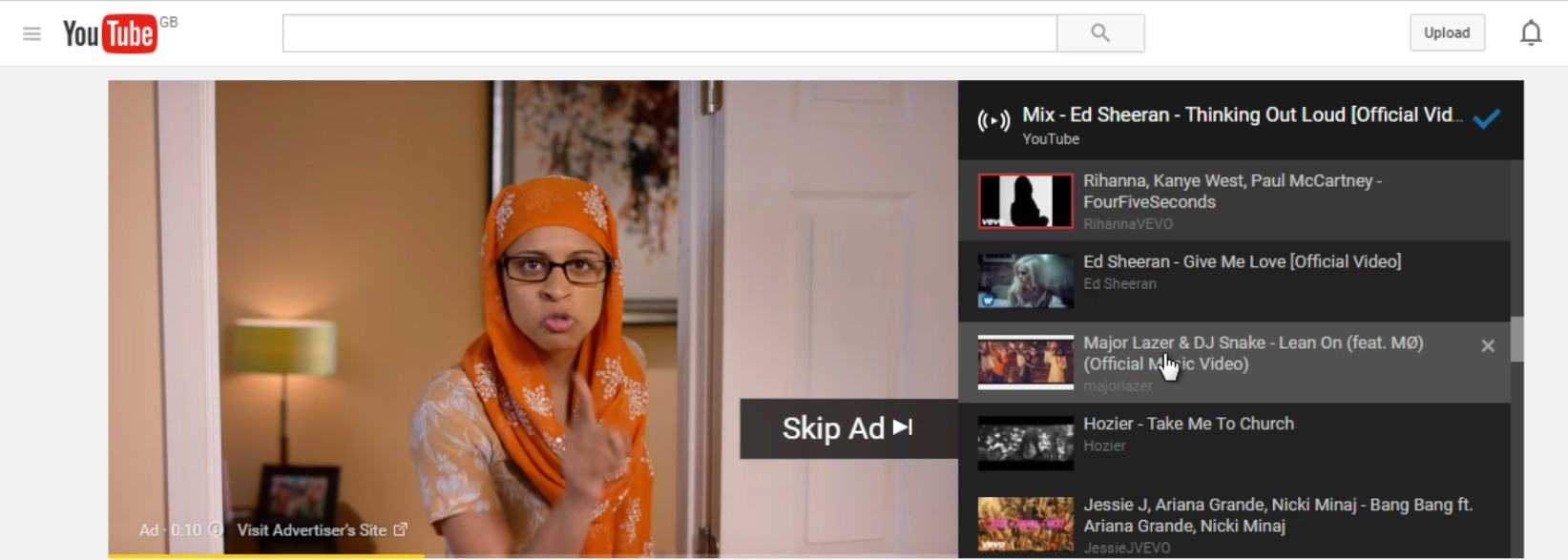}}\vspace{-8pt}
\end{figure}

\vspace{-6pt}\textbf{4. Skip YouTube ads} (Monitoring) is a standby task that clicks the text \textit{Skip Ad} if/when it appears during a YouTube video. This task illustrates the need for spontaneous responses, because the \textit{Skip Ad} advertisement banners appear randomly, for varying periods of time, ranging from 10 seconds to a few minutes, during playback of the requested content. Figure~\ref{fig:scene4} demonstrates an example of the \textit{Skip ad} task.

\begin{figure}[h]
%\floatbox[{\capbeside\thisfloatsetup{capbesideposition={right,top},capbesidewidth=4cm}}]{figure}[\FBwidth]
{\caption{Close-ups of YouTube Ads. These ads appear at the bottom of a playing video, and our system detects and successfully closes them in Scenario 5.}\label{fig:closeadv}}
{\includegraphics[height=2.5cm]{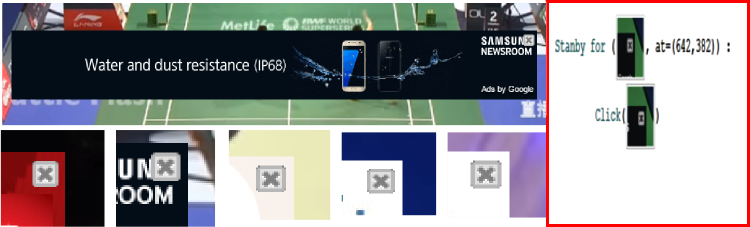}}\vspace{-8pt}
\end{figure}

\vspace{-6pt}\textbf{5. Close YouTube ads} (Monitoring) creates a standby script to close advertisements that may appear, despite various changing backgrounds, as shown in Figure~\ref{fig:closeadv}. The first line of the script directs the system to monitor an area where the given pattern can appear. When the system detects the pattern, the system triggers a script, in the second line, to click on that pattern.

\textbf{Looping:} Our system allows a loop to be a step in a
linear action, or to be a stand-alone script. Looping tasks are the tasks that execute the same sequences of linear actions multiple times on similar looking yet different objects.

To indicate the start and stop of a loop, the demonstrator inputs the looping signal key combination Ctrl+Shift+l or Ctrl+Shift+Break, before and after performing one sequence of actions that need to be repeated. Thereafter, the instructor gives examples of patterns that needed to be a starting point of the loop by pressing a Ctrl key while clicking on an example pattern. When the instructor is happy with the examples, they then input the looping signal once more. %to indicate the end of looping.
The script can be followed by linear actions or can finish right after the third looping signal. % by pressing the end of sequences keys combination, Shift+Esc. 

The teaching phase of a looping task has one additional step. The system displays the result of the trained Random Forest, and lets the teacher add positive examples, remove false positive results and provide supporters. This triggers re-training.

\begin{figure}[h]
%\floatbox[{\capbeside\thisfloatsetup{capbesideposition={right,top},capbesidewidth=4cm}}]{figure}[\FBwidth]
{\caption{Scenario 6: create slides from folder full of images. The generated script is shown in the red frame. The system starts by building a list of locations that will be the starting points for each iteration. The list is formed by the Trained RF, which trained and refined in the teaching phase with a few examples stemming from the demonstration phase. The system then iteratively executes a sequence of actions from line three to five (DragTo, Click, Click). In this scenario, the two applications are displayed side-by-side.}\label{fig:scene6}}
{\includegraphics[height=4.7cm]{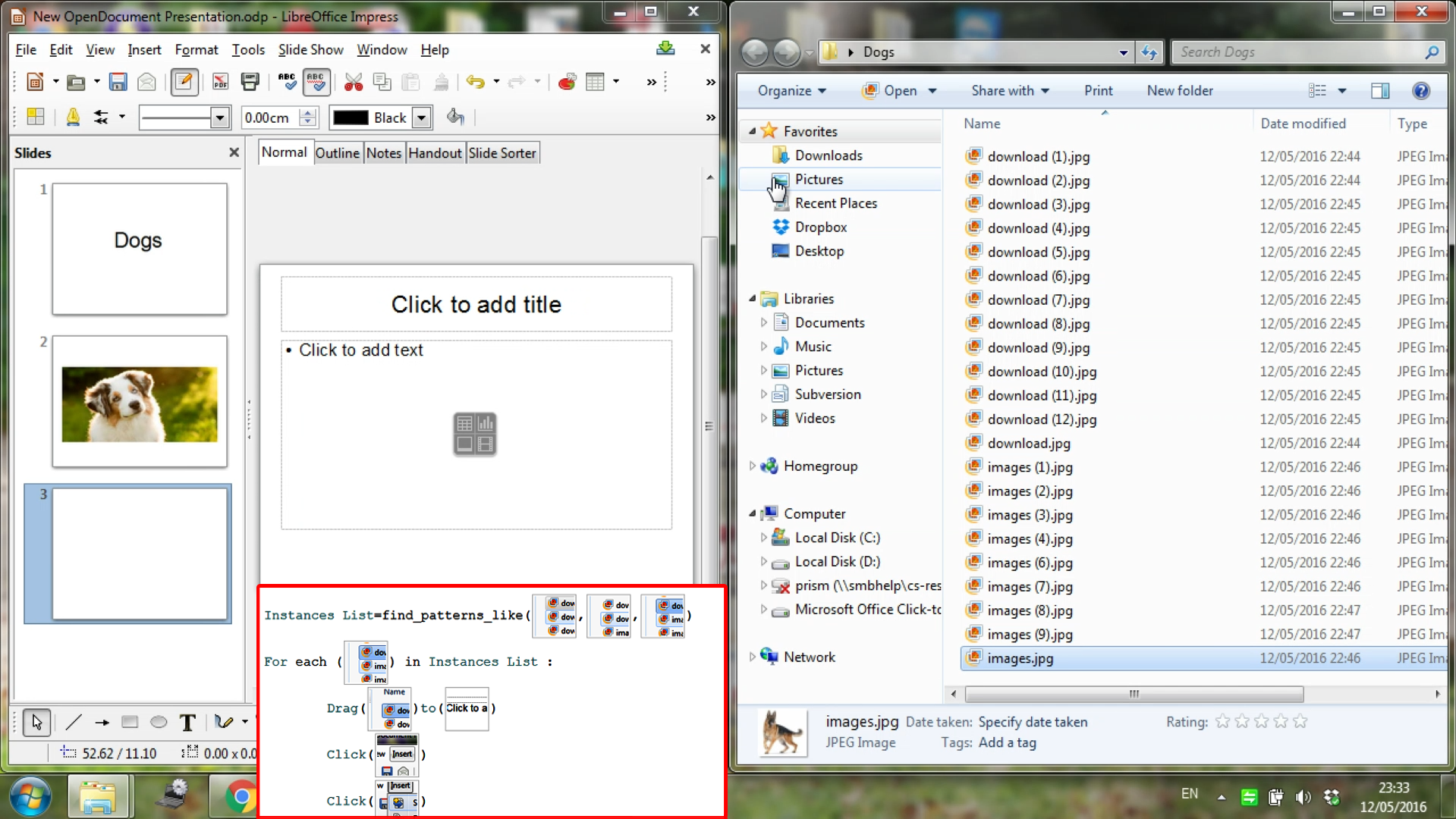}}\vspace{-8pt}
\end{figure}

\vspace{-6pt}\textbf{6. Create slides out of jpgs folder} (Looping) The task is to create a presentation where each slide features one image from a given folder. To create the script, a demonstrator only shows how to create one slide from one jpg, and gives a few examples of what the jpg file-icon looks like. In the running phase, the system steps through all the jpg files in a given folder, making each one into a separate slide of the LibreOffice Impress presentation. Not only does this show that the system can loop, the task also demonstrates that the system can help the user complete repeated steps across different applications (LibreOffice Impress and Windows Explorer). We show an example screen of the task and the generated script in Figure~\ref{fig:scene6}.

\begin{figure}[h]
%\floatbox[{\capbeside\thisfloatsetup{capbesideposition={right,top},capbesidewidth=4cm}}]{figure}[\FBwidth]
{\caption{Two application's screens from Scenario 7, where file names are being collected into a spreadsheet. The script of the task, in the red frame, involves switching back and forth between the two applications, and pasting the text into similar-looking cells.}\label{fig:scene7}}
{\includegraphics[height=9cm]{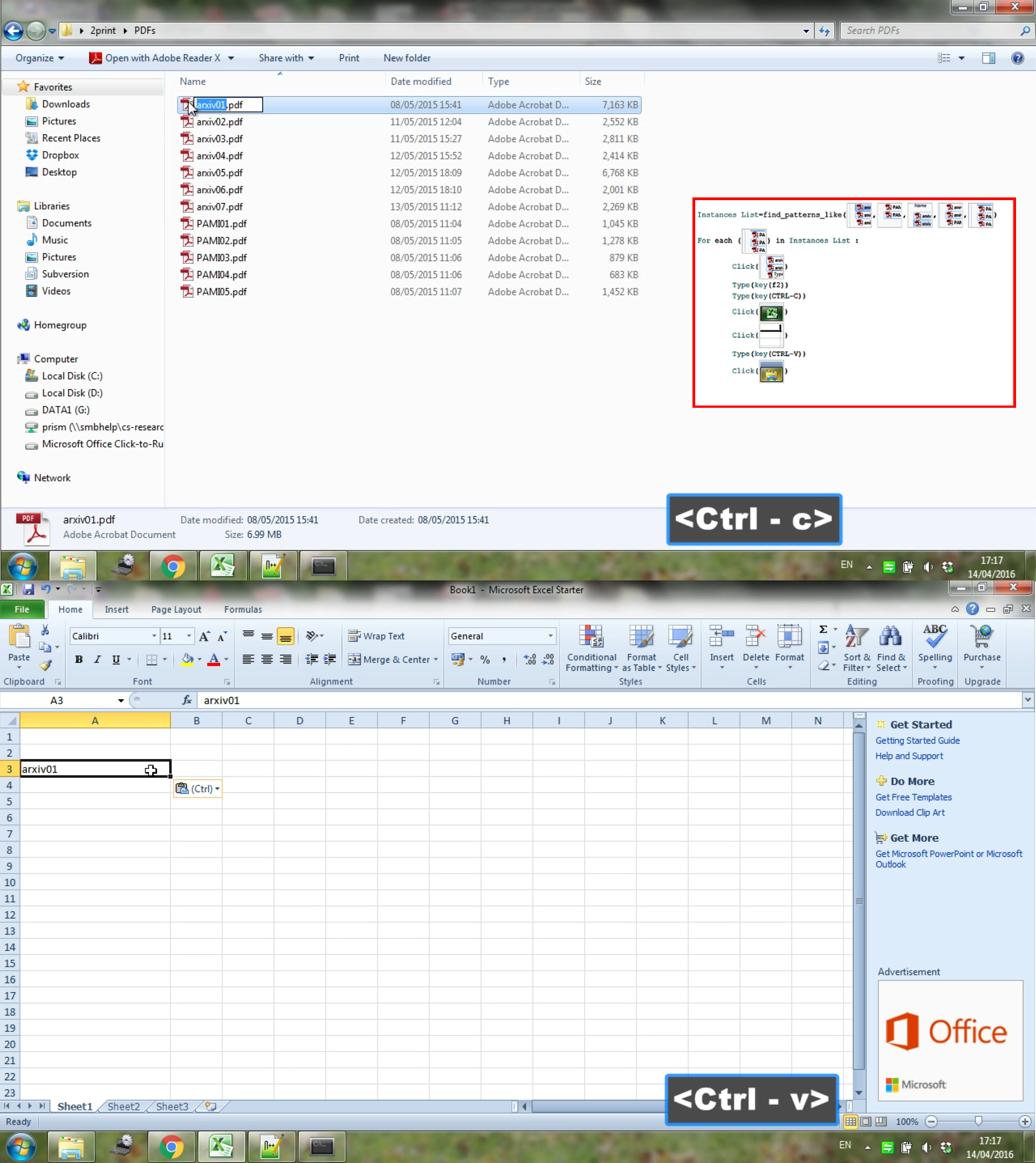}}\vspace{-8pt}
\end{figure}

\vspace{-6pt}\textbf{7. Create spreadsheet of filenames} (Looping) The purpose of this scenario is to create a list of filenames in a spreadsheet program, Figure~\ref{fig:scene7}. In the running phase, the system copies the filenames from within a given folder into successive Microsoft Excel cells. While repeatedly successful, the paste operation targeted the cell below the previously-selected (dark outline) cell on the spreadsheet. So the first entry will always be pasted below the initially selected cell.

\begin{figure}[h]
%\floatbox[{\capbeside\thisfloatsetup{capbesideposition={right,top},capbesidewidth=4cm}}]{figure}[\FBwidth]
{\includegraphics[height=9cm]{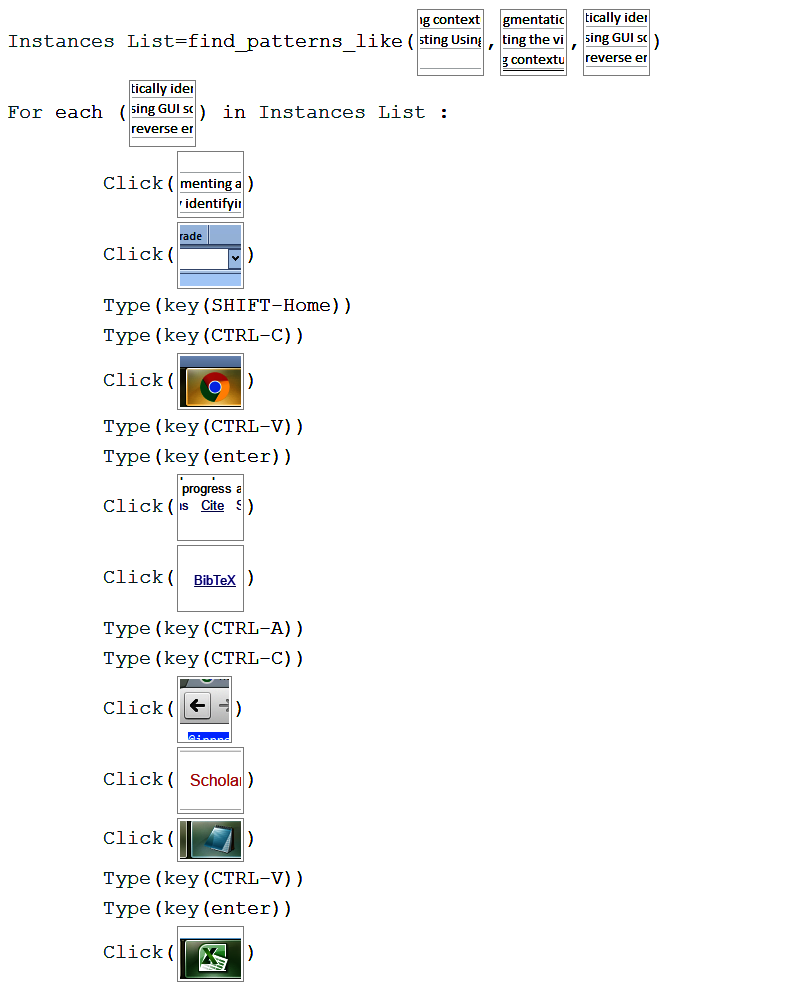}}\vspace{-8pt}
{\caption{A synthesized script of Scenario 8, where a BibTex file is automatically constructed from a list of paper titles. Three different desktop GUI's were involved. The user was able to train the system quite easily, and can just ran the task without further instructions when writing their next research paper.}\label{fig:scene8}}
\end{figure}

\vspace{-6pt}\textbf{8. Create BibTex from spreadsheet} (Looping) This is the most complex of all the scenarios listed here. The task involves switching between three different applications (eight different screens). An instructors needs to plan out the task's steps, to ensure each application is in a state that is ready for the same action of the next loop to be executed. In the running phase, the system works through an Excel file that lists titles of papers that should be cited. The system then uses Google Scholar to look for the BibTex of each paper, and produces a single BibTex file listing all the citations using the Notepad program. Figure~\ref{fig:scene8} illustrates the generated script.

\vspace{-6pt}\textbf{9. Move MSWord files to a folder} (Looping-Video) In this scenario, we tested a further proof of concept of our system. The system successfully uses \emph{only} video from a screencast as input, instead of data from the sniffer, illustrating that instructors could post how-to-videos online, which can then easily be refined into a working script. 

\begin{figure}
%\floatbox[{\capbeside\thisfloatsetup{capbesideposition={right,top},capbesidewidth=4cm}}]{figure}[\FBwidth]
{\includegraphics[height=9cm]{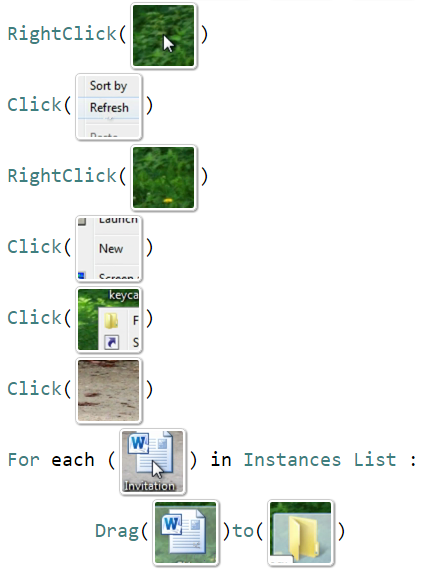}}
{\caption{The system successfully use videos as input of the system, instead of generating the input log-file from the sniffer, to create a working script. Please note that the system failed to remove mouse pointer from the target patterns in the first and the seventh lines.}\label{fig:scene9}}
\end{figure}

The scenario starts by executing a sequence of linear basic actions to create a new folder. It then continues to iteratively Drag and Drop each Microsoft Word file into the newly created folder. The script is shown in Figure~\ref{fig:scene9}.

\begin{table*}
\centering
{
\small
\begin{tabular}{|c|c|c|c|c|c|c|c|c|c|c|c|}
\cline{2-6}
 \multicolumn{1}{l|}{} & \multicolumn{5}{c|}{Basic Actions + Typing}& \multicolumn{3}{l}{} \\ \hline
\begin{tabular}[r]{@{}l@{}}\rotatebox{90}{Scenario}\end{tabular} & \begin{tabular}[r]{@{}l@{}}\rotatebox{90}{Click}\end{tabular}& \begin{tabular}[r]{@{}l@{}}\rotatebox{90}{Click Drag}\end{tabular} & \begin{tabular}[r]{@{}l@{}}\rotatebox{90}{Double Click \hspace{0.1cm}}\end{tabular} & \begin{tabular}[r]{@{}l@{}}\rotatebox{90}{Right Click}\end{tabular} & \begin{tabular}[r]{@{}l@{}}\rotatebox{90}{Typing}\end{tabular} &  \multicolumn{2}{l|}{\begin{tabular}[l]{@{}c@{}}Transcription \\ \end{tabular}} &  \multicolumn{2}{l|}{\begin{tabular}[l]{@{}c@{}}Reproduction \\ \end{tabular}} & \multicolumn{2}{l|}{\begin{tabular}[c]{@{}c}Training Time \\  VS \\Refining Time (average) \\ \end{tabular}} \\ \cline{7-12}
 & & & & & & \begin{tabular}[l]{@{}c@{}}Sikuli\\Slides \end{tabular} & Our & \begin{tabular}[l]{@{}c@{}}Sikuli\\Slides \end{tabular} & Our & \begin{tabular}[l]{@{}c@{}}Sikuli\\Slides \end{tabular} & Our\\ \hline
1 (Linear)  &   2    &   0   & 0 & 0 & 0 & \cmark  & \cmark & \cmark * & \cmark &  10s/49s  &    10s/27s    \\ \hline
2  (Linear) &   6    &   1   & 0 & 0 & 0 & \cmark * & \cmark & \cmark ** & \cmark &  27s/10m  &   27s/170s    \\ \hline
3  (Linear)  &   11    &   0   & 0 & 0 & 4 & \cmark * & \cmark & \xmark & \cmark  &    40s/$\infty$    &   40s/4m        \\ \hline
3.2   (Linear)  &   13    &   0   & 0 & 4 & 0 & \cmark * & \cmark & \xmark & \cmark  &    37s/$\infty$     &   37s/7m     \\ \hline
4. (Monitoring)   &   1    &   0   & 0 & 0 & 0 & \xmark & \cmark & \xmark & \cmark & N/A &    10s/5.5m    \\ \hline
5. (Monitoring)  &   1    &   0   & 0 & 0 & 0 & \xmark & \cmark & \xmark & \cmark & N/A   &    12s/6.9m    \\ \hline
6. (Looping)  &   2x    &   1x   & 0 & 0 & 0 & \xmark & \cmark & \xmark & \cmark & N/A   &    35s/10m   \\ \hline
7. (Looping)  &   4x    &   2x   & 0 & 0 & 4x & \xmark & \cmark & \xmark & \cmark  & N/A    &   60s/6.6m     \\ \hline
8. (Looping)  &   9x    &   0   & 0 & 0 & 8x  & \xmark & \cmark & \xmark & \cmark  & N/A   &   86s/12.5m     \\ \hline
9. (Looping-Video)  &   4    &   1x   & 0 & 2 & 0  & \xmark & \cmark & \xmark & \cmark  & N/A   &   25s/\textbf{22m}     \\ \hline
\end{tabular}
}%\vspace{-10pt}
\caption{User study on our system compared to Sikuli Slides. Scenario 3.2 is an alternative way to perform Scenario 3, without pressing shortcut key combinations that Sikuli Slides is known to be missing. Nevertheless, we eventually realized that Sikuli Slides isn't detecting the right click actions either. (\cmark = successful, \cmark * = partially successful, \cmark ** = can be successful with guidance from the operator, \xmark = can not succeed at the task at all). x represents the number of repeated loops needed to complete the task. Please note that $90\%$ of the refining time for Task 9 is offline - devoted to the time spent on processing video to produce the log-file.}
\label{tab:result}
\end{table*}

\section{Discussion and Future Work}\label{sec:conclusion}

The two sets of evaluation scenarios showed that our approach substantially extends the programming by demonstration functionality that was available to non-programming users of desktop-automation tools. The main innovation is the sanity-check performed when the instructor demonstrates their task: given a cooperative human, it allows the system to transition from a winner-takes-all template-matching view of targets and actions, onto a supervised-classification interpretation of the instructor's intentions.

This prototype has important opportunities for improvements. Basic actions are occasionally misclassified, when none of them has a high probability. %, and this required the instructor to re-demonstrate in up to 1 of 5 cases. 
Tests showed the joint segmentation and classification algorithm has an average accuracy of $95.4\%$ for classifying each basic action. Our system allows users to fix misclassified actions instead of requiring a user to re-record the instruction again. Users were more successful and could do more with our system, but found the concept of supporters somewhat foreign, at least as presented in our instructions. 

Currently, the system works without the awareness of states of the computer. For example, if a task expects to work with a pre-opened folder (or to open a closed one), the end-user must prepare their desktop appropriately. In addition, we insert short fixed-length sleep() after each action to account for loading time of the computer because the system cannot know if the OS task has finished/ web-page has loaded. Therefore, shorter sleeps would make automated tasks go faster, but could ask for actions before the GUI is ready. This could be addressed in the future by train the system to recognize computer states from visual signals.  
%Currently, there is no robustness to the initial state of the computer: if a task expects to work with a pre-opened folder (or to open a closed one), the end-user must prepare their desktop appropriately. This could be addressed in the future by intelligently chaining tasks together, and showing temporary pop-up messages to indicate success or failure of a task.
% Also talk about lists of things, and how the middle looks different from the top and bottom.
Finally, the current appearance models have fixed size and aspect ratio, which can hurt accuracy when items in a list are short and wide. Learned appearance features, even spanning across devices, could emerge, given enough training footage.

\section{Acknowledgments}

We thank all the volunteers, and all reviewers, who provided helpful comments on previous versions of this paper. Authors gratefully acknowledge
the Ministry of Science and Technology of Thailand Scholarship and the EPSRC grants EP/K023578/1 and EP/K015664/1.

\clearpage
\bibliographystyle{SIGCHI-Reference-Format}
\bibliography{IUI}
\end{document}